\documentclass[]{elsart}
\usepackage{graphics}
\usepackage{graphicx}
\usepackage{epsfig}
\usepackage{amssymb}
\begin{document}
\begin{frontmatter}
\title{The mean-field approximation model of company's income growth}


\author{\small{Takayuki Mizuno$^1$$^{a}$}},
\author{\small{Misako Takayasu$^{b}$}}, and 
\author{\small{Hideki Takayasu$^{c}$}}

\address{$^{a}$Department of Physics, Faculty of Science and Engineering, Chuo University, 1-13-27 Kasuga, Bunkyo-ku, Tokyo 112-8551, Japan}
\address{$^{b}$Department of Complex Systems, Future University-Hakodate, 116-2 Kameda-Nakano-cho, Hakodate, Hokkaido 041-8655, Japan}
\address{$^{c}$Sony Computer Science Laboratories Inc., 3-14-13 Higashigotanda, Shinagawa-ku, Tokyo 141-0022, Japan}

\thanks{Corresponding author.\\
{\it E-mail address:}\/ mizuno@phys.chuo-u.ac.jp (T.Mizuno)}

\begin{abstract}
We introduce a mean-field type approximation for description of company's 
income statistics. Utilizing huge company data we show that a discrete 
version of Langevin equation with additive and multiplicative noises can 
appropriately describe the time evolution of a company's income fluctuation 
in statistical sense. The Zipf's law of income distribution is shown to be 
hold in a steady-sate widely, and country-dependence of income distribution 
can also be nicely implemented in our numerical simulation.
\end{abstract}

\begin{keyword}
Econophysics \sep Company's Income \sep Zip's Law
\PACS 89.65.Gh,02.50,05.40.+j
\end{keyword}

\end{frontmatter}

\section{Introduction}
Studies on wealth distribution can be traced back more than 100 years. It is 
widely known that Pareto published the first report on the individual income 
distribution in 1897 [1]. He showed that the probability density 
distribution of income follows a power law distribution in the high-income 
range. For smaller income the cumulative probability distribution in log-log 
plot clearly deviates from the power law. This range of individual income is 
reported to be approximated by a log normal distribution [2] or an 
exponential distribution [3,4,5]. In recent years huge data bases of income 
data are accessible to anyone due to the progress of computer technology, 
and a new type of research focusing on macroscopic statistical laws deriving 
from microscopic economics data is prosperous [6,7]. 

Similar analysis of income statistics has been done also for companies. 
Company's data is finely described more than the individual because the 
accounting of the company is directly related to the tax revenues and the 
information is essential for investors. It is well-known that the size 
distribution of companies is also characterized by a power law distribution 
[8]. Distribution of sales follows a log-normal distribution [9] and the 
growth of sales shows sporadic violent fluctuations [10,11]. In the case of 
Japanese companies income the cumulative probability distribution clearly 
follows a power law with exponent --1 so-called the Zip's law [12]. The Zipf 
law of company statistics is also known about the distribution of the number 
of employees [13]. 

The wealth distribution among individual investors is theoretically studied 
using a generalized Lotka-Volterra model [14,15] and the relation between 
the exponent of wealth distribution and the number of agents in the market 
is discussed in detail [16]. It is pointed out that a multiplicative process 
plays an essential role in the dynamics of wealth growth. There are other 
types of models that realize the distribution of individual income by 
introducing exchanges of wealth between agents [17,18]. Similar approach was 
already introduced in the study of company size distribution by modeling the 
competition among companies [19]. 

It is reported that Zipf's law of Japanese income distribution has been 
maintained over 30 years with very high precision [6,20]. This is a special 
feature compared with the case of personal income distribution [21] in which 
the exponent apparently changes year by year. It is likely that the income 
statistics of companies can be regarded as a steady stochastic process 
maintaining the power law distribution. In the study of statistical physics 
it is known that one of the simplest stochastic processes producing a steady 
power law distribution is the discrete version of Langevin equation with 
additive and multiplicative noises [22, 23]. A preliminary study on modeling 
company's income statistics based on such type of Langevin equation has 
produced promising results explaining the basic power law distribution from 
the statistics of growth rates [6].

In this paper we analyze databases of companies' income of three countries 
following this formulation. We introduce a kind of mean-field type evolution 
equation of a company's annual income that involves a multiplicative growth 
rate and an additive noise term, having the identical form with the discrete 
Langevin equation. Precise parameter fittings can be done based on this 
modeling and our numerical simulations can reproduce the present income 
distributions for each country. With this technique we now understand the 
meaning of the Zipf's law and also we can predict the future development of 
company income distribution, implying that this new statistical physics 
approach can bridge econophysics and the macro-economic theory.

\section{Income Distribution}
At first we investigate statistical laws of company's income by analyzing 
the data ( about 15,000 companies all over the world except USA [24], about 
15,000 companies in USA [25], and Japanese companies whose annual incomes 
exceed 40 million yen, that is, about 80,000 companies [26] ). Here, income 
is defined by the total incoming cash flows minus outgoing ones before 
taxation, therefore, it can take a negative value although major companies' 
incomes are nearly always positive. 

Incomes of each company show large fluctuations year by year as shown in 
Fig.1. It is not rare that this year's income is increased tenfold or 
reduced to one-tenth from last year's income. Namely, it is an amazing fact 
that the total distribution of income distribution is keeping the identical 
power law for more than 30 years as each component company's income changes 
significantly and even there are many companies disappeared during this 
period. Power law distribution is also observed in case of USA and other 
countries. In Fig.2 we plot the distributions of income in Japan, UK, and 
USA, and in each case the distribution is confirmed to be close to a power 
law with exponent -1. For USA we also plotted the distribution of negative 
incomes by their absolute values for comparison. It is a non-trivial fact 
that the negative income distribution also follows the similar distribution 
having a power law tail [3].

\section{The Mean-Field Approximation Model of Income Growth}
 As mentioned above we assume the following form of stochastic time 
evolution of income $I$ for each company,

\begin{equation}
\label{eq1}
I(t + 1) = \alpha (t) \cdot B(t:I) \cdot I(t) + f(t),
\end{equation}

\noindent
where, $B(t:I)$ is a multiplicative noise representing the income growth, 
$f(t)$ is the additive noise, and the coefficient $\alpha (t)$ specifies the 
sign of income, namely, it takes either 1 or -1 with empirically determined 
probability. It is known that the steady power law realizes if the 
multiplicative factor $B(t:I)$ and the additive noise $f(t)$ randomly 
fluctuate independently with respect to $I(t)$[22]. By introducing the 
coefficient $\alpha (t)$ we can take into account the cases with negative 
incomes when a company slumps suddenly. We find the occurrence of such cases 
even for very large companies with probability about 3{\%}.

\section{The Growth coefficient $B(t:I)$ in Model}
We estimate the growth coefficient $B(t:I)$ by investigating the 
distribution of $\left( {R(t) \equiv } \right)I(t + 1) / I(t)$ in the range 
of large $I(t)$ because $f(t) / I(t)$ is expected to be very small in this 
range. In Fig.3 we show the growth rate distributions of income from 1989 up 
to 1995 in USA. It is found from our data analysis that the coefficient 
$B(t:I)$ is dependent on income $I(t)$. In order to take into account this 
size dependence we normalize the growth rate of income by $\sigma (I) / 
\sigma _0 $, where $\sigma (I)$ is the standard deviation of logarithmic 
growth rates $\log \left( {I(t + 1) / I(t)} \right)$ and $\sigma _0 $ is the 
standard deviation observed in the large income range. The curves of 
distribution of normalized growth rates $R'(t)$ for different income categories 
collapse nicely to a curve as shown in Fig.4.

\section{Coefficients $\alpha (t)$ and $f(t)$ in Model}
For estimation of the additive noise term $f(t)$ we can utilize the sign-change 
probability of income in the following way. As shown in Fig.5 the observed 
sign-change probability is nearly constant for large $\vert I(t)\vert $, while 
its value increases for smaller $\vert I(t)\vert $. This empirical fact can be 
approximated by the simple assumption that the statistics of $f(t)$ is 
independent of $I(t)$ and its standard deviation is given by the value of $I$ at 
which the sign-change probability begins to bend, that is, about 400 in the 
case of Fig.5. Actually this model gives the sign-change probability as 
shown by the dotted curves in Fig.5. 

Now our basic model equation is given as

\begin{equation}
\label{eq2}
I(t + 1) = \alpha (t) \cdot R'(t)^{\frac{\sigma (I)}{\sigma _0 }} \cdot I(t) 
+ f(t) .
\end{equation}

\noindent
where the function of $\alpha (t)$ for USA companies, for example, is given by 

\begin{equation}
\alpha (t) = 1 with probability  0.97 (I(t) > 0), 0.75 (I(t) < 0)
\end{equation}
\hspace{1.5cm} $= - 1 with probability 0.03 (I(t) > 0), 0.25 (I(t) < 0) .$

\section{Monte Carlo Simulations and Theoretical Analysis}
We now perform Monte Carlo simulation of the growth of income of each 
company using the values of coefficients estimated from the real data for 
each country. The initial value of income of any company used in our 
simulation is 100 and the number of companies is 60000. The time development 
of distributions of both positive and negative incomes in the case of USA is 
shown in Fig.6. The simulation result with $t=50$ (meaning 50 years) matches 
the actual present distribution nicely. Results for income distributions for 
UK and Japan are plotted in Fig.7 together with the results of t=50 for USA. 
In the case of Japan a steady-state distribution realizes at about 25 years 
and in the case of UK the best fit curve is obtained for $t=100$ by comparing 
with Fig.2. In Japanese case it is confirmed that growth coefficient 
distribution does not depend on the income range, namely, the 
coefficient $\sigma(I)$ of Eq.(\ref{eq2}) is treated as a constant and the mean 
value of the growth rate is very close to 1 [6]. It is proved theoretically 
that in such case the income distribution at the steady state is 
characterized by a power law with exponent -1 by applying the formulas of 
random multiplicative Langevin equation relating the growth rate 
distribution and the steady-state distribution [22]. 

\begin{equation}
 < B(t:I)^\beta > = 1 \quad ,
\quad
P(I) \propto I^{ - \beta } \quad ,with \beta = 1
\end{equation}

This theoretical result is consistent with the fact that the observed power 
exponent is about -1 for over 30 years in Japan.

\section{Discussion}
We can expect future income distribution by assuming that future income 
growth rate distribution to be the same as that of past income growth 
because this model is time evolution equation. For example, the company's 
income distribution in USA is expected to be still growing, namely, assuming 
that the present company growth rate distribution being kept identical then 
our simulation result predicts that the income distribution will keep 
growing for more than 100 years. 

For practical purposes our model may contribute to a policy of tax revenues 
or evaluation of investment strategy. It is possible to estimate the 
statistical outcome of investment for a group of companies, actually our 
preliminary simulation results suggests that investment for companies with 
intermediate income is most profitable with intermediate risks. 

As a new field of physics our approach paves the way to bridge 
macro-economics and micro-economics by the concepts and methods developed in 
statistical physics based on actual data. It may be impossible even in the 
future to describe fully the activity of a company in terminology of 
physics, however, the overall statistical properties can be nicely 
approximated by a rather simple stochastic equation familiar in physics as 
demonstrated in this paper. We believe this new field will be promising for 
development of both physics and economics.

\textbf{Acknowledgments}

We would like to thank Mr. Kenichi Tsuboi and Ms. Nobue Sakakibara of 
Diamond Publishing Co. Ltd. for allowing us to use the data for Japanese 
companies. Also, we appreciate H.E.Stanely, M. Katori and S. Kurihara for 
stimulus discussions.




\newpage 
\begin{figure}
\begin{center}
\epsfig{file=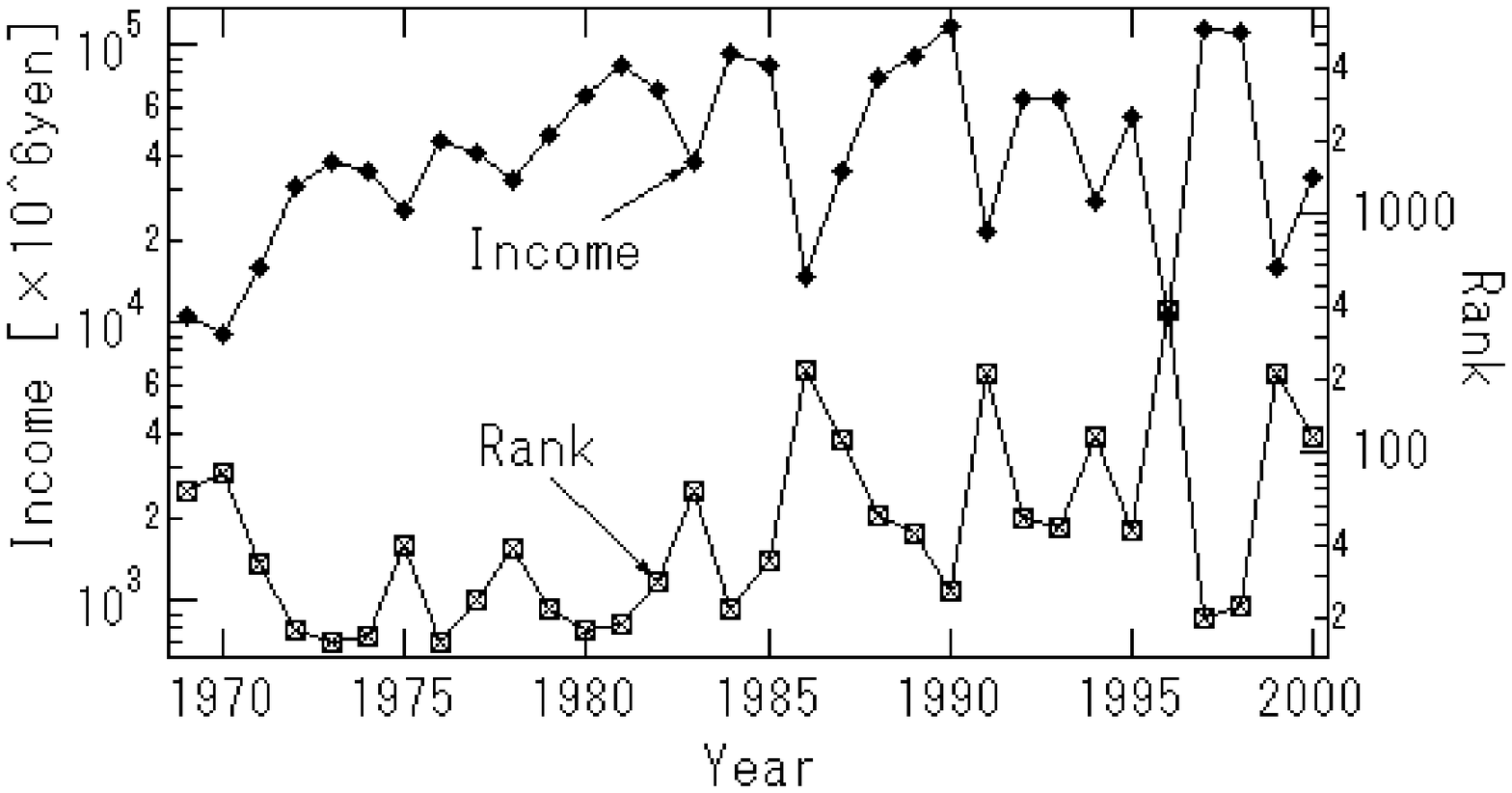,width=8cm}
\end{center}
\caption{An example of income fluctuation of a Japanese company for 30 years in 
log-scale. White and dark squares show incomes and corresponding rankings, 
respectively.}
\end{figure}

\begin{figure}
\begin{center}
\epsfig{file=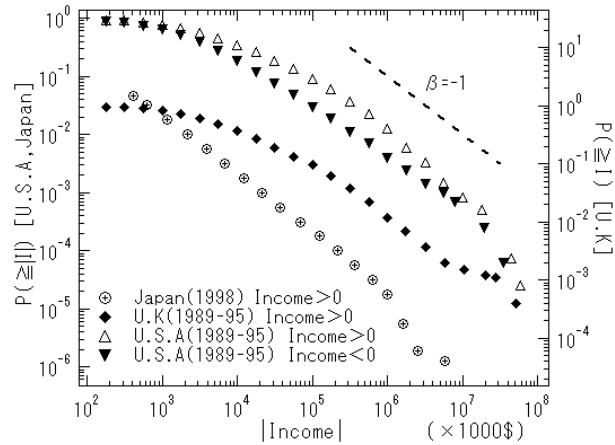,width=8cm}
\end{center}
\caption{Log-log plot of cumulative distributions of positive 
company's incomes in Japan, UK, USA and negative income in USA. 
The power exponent of the dashed line is -1. 
In the case of UK the distribution for low income is distorted due to lack 
of data.}
\end{figure}

\begin{figure}
\begin{center}
\epsfig{file=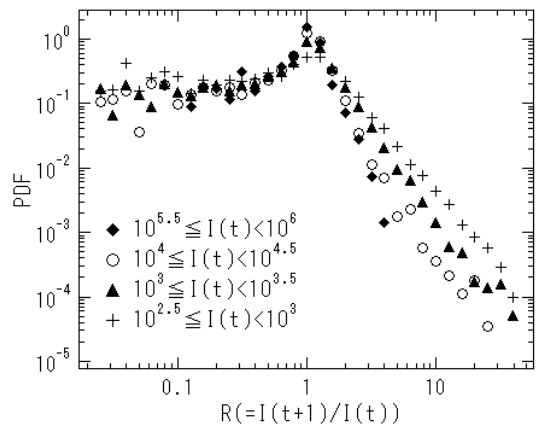,width=7cm}
\end{center}
\caption{Log-log plot of the probability density of income growth 
rates of USA companies. 
The unit of income $I(t)$ in the figure is 1000{\$}.}
\end{figure}

\newpage 

\begin{figure}
\begin{center}
\epsfig{file=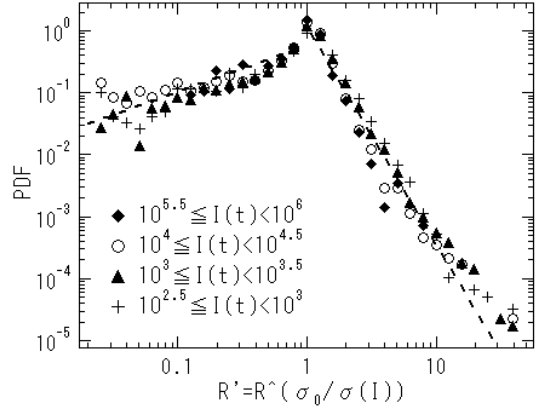,width=7cm}
\end{center}
\caption{Normalized income growth rate distribution of USA companies. 
The dashed lines show approximation by power laws used in the numerical 
simulation.}
\end{figure}

\begin{figure}
\begin{center}
\epsfig{file=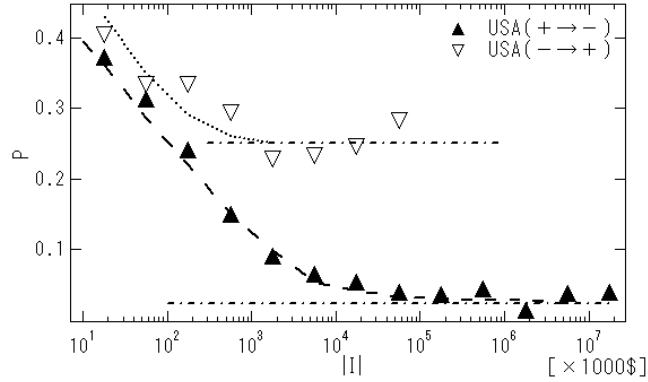,width=8.5cm}
\end{center}
\caption{The sign change probability of income as a function of income 
size. 
Dark triangles show sign change probability from positive income to negative 
income, and white triangles show the probability of opposite cases.}
\end{figure}

\newpage 

\begin{figure}
\begin{center}
\epsfig{file=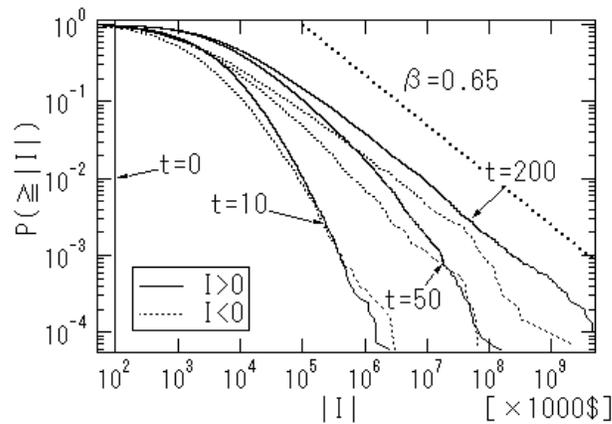,width=8cm}
\end{center}
\caption{Simulation results of income growth for USA. 
The line of USA for $t=50$ agrees with the actual distribution in Fig.2. }
\end{figure}

\begin{figure}
\begin{center}
\epsfig{file=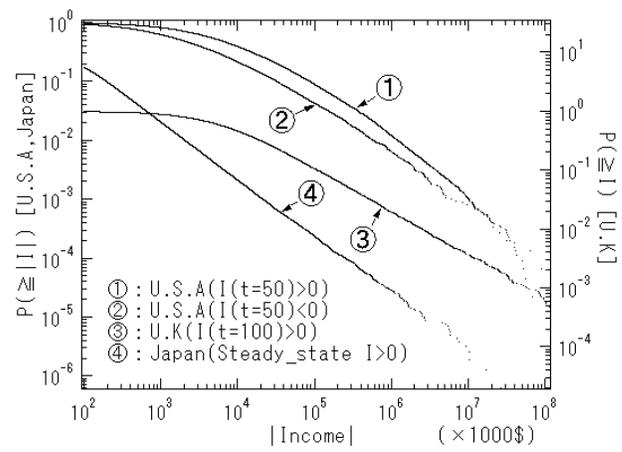,width=8cm}
\end{center}
\caption{The simulation results for Japan, UK and USA.}
\end{figure}

\end{document}